\documentclass[12pt]{article}
\usepackage[dvips]{graphicx}
\renewcommand {\d}  {\hbox{d\hskip-1.1ex{\raise0.640ex\hbox{--}}
\skip 0.70ex}}
\newcommand   {\D}  {\hbox{D\hskip-1.9ex{\raise0.175ex\hbox{--}}
\hskip0.85ex}}

\begin{document}

\newcounter{buco}
\setcounter{equation}{0}
\setcounter{buco}{0}
\renewcommand{\thesection}{\arabic{section}}
\renewcommand{\theequation}{\arabic{section}\arabic{equation}}

% \setlength{\voffset}{-0.5in} \setlength{\textheight}{8.5in}
%\def\baselinestretch{2}
%\@addtoreset{equation}{section}
\def\theequation{\thesection.\arabic{equation}}
\def\theeqnarray{\thesection.\arabic{eqnarray}}

\begin{center}
\begin{Large}
\begin{bf}
Two photon decays of scalar mesons in a covariant quark model\\
\end{bf}
\end{Large}
\vspace{2.0cm}
S. Fajfer\footnote{E-mail: Svjetlana.Fajfer@ijs.si} \\
J. Stefan Institute, Jamova 39, P. O. Box 3000, 1001 Ljubljana,
Slovenia

\vspace{2.0cm}
D. Horvati\'{c}, D. Tadi\' c \footnote{E-mail: tadic@phy.hr} and  S. \v Zganec
%%@
\\
Physics Department, University of Zagreb, Bijeni\v cka c. 32, 10000 Zagreb,
Croatia

\end{center}
\vskip 2cm
%%%%%%%%%%%%%%%%%%%%%%%%%%%%%%%%%%%%%%%%%%%%%%%%%%%%%%%%%%%%%%%%%%%%%%%%%
%%%%%%%%%%%%%%%%%%%%% ABSTRACT  %%%%%%%%%%%%%%%%%%%%%%%%%%%%%%%%%%%%%%%%%
%%%%%%%%%%%%%%%%%%%%%%%%%%%%%%%%%%%%%%%%%%%%%%%%%%%%%%%%%%%%%%%%%%%%%%%%%

\begin{center}
\bf Abstract
\end{center}
\baselineskip=24pt

\noindent
Two photon decay widths of the $J^P = O^+$ scalar mesons
$a_{0} (980)$, $f_{0}(980)$,
$f_{0}(1370)$ and $\chi_{c0}$ are calculated in a covariant
model which is
characterized by the quark - antiquark structure.
Previously such models were used to calculate current form factors.
Here a different application is tried. A simple version of the model uses
adjusted nonrelativistic model parameters with small quark masses. The results
%%@
seem to prefer nonideal mixing
of $f_0(980)$ and  $f_0(1370)$.
The calculated decay rate of $\chi_{c0}$ agrees with the experimental results.

\bigskip
{\it{Short title}}: Two photon decays of scalar mesons \\


\newpage
%%%%%%%%%%%%%%%%%%%%%%%%%%%%%%%%%%%%%%%%%%%%%%%%%%%%%%%%%%%%%%%%%%%%%%%%%
%%%%%%%%%%%%%%%%%%%%%%%%% SECTION 1 %%%%%%%%%%%%%%%%%%%%%%%%%%%%%%%%%%%%%
\vspace*{1pt}   %) USE THIS MEASUREMENT WHEN THERE IS
\section{Introduction}  %) A SECTION HEADING
\vspace*{-0.5pt} \noindent The scalar mesons appearing around 1
$GeV$ mass scale seems to be least understood among spin zero
mesons. The experimental data\cite{1}, on strong decays of
$a_{0}(980)$, $f_{0}(980)$ and $f_{0}(1370)$ do not lead to the
final understanding of the real structure of these mesons
\cite{2}$-$\cite{PLB98}.
Ideas exists that these states are $K \bar K$ molecules \cite{2}$-$\cite{7}.
There is a %%@
suggestion that $a_0(980)$ has a four-quark structure with the
strange quarks ($s \bar s$) contribution\cite{Achasov}, based on
the  analyses of the $\phi \to \gamma a_0(980) \to \gamma \eta
\pi$\cite{PLB98}. Such uncertainties suggest further
investigation, which were  concentrated on $2 \gamma$ decays of
$a_{0} (980)$, $f_{0}(980)$, $f_{0}(1370)$ and $\chi_{c0}$ mesons.
The aim was to explain the experimental decay widths and at the
same time to reproduce the measured masses.

The relativistic  quark model is used to correlate various data
and to establish the connection between masses and decay widths.

For that purpose one employs the covariant model \cite{16,18}
which includes the heavy - quark symmetry. As shown in the
following section the model and the calculations are both
covariant and gauge invariant.
This model is a covariant generalization \cite{16,18} of the well known ISGW %%@
model \cite{17}. However, here the usage of small quark masses is
investigated, which
means the avoidance of the weak binding limit approximation in its strictest %%@
sense \cite{17}.
That might better mimick the real quark fields which should appear in the %%@
photon emitting quark loop (Fig. 1) in the first order of QED/QCD
expansion. In a
very simplified version of that model, which is employed here, only the quark
%%@
momentum distribution parameter $\beta$ and the model quark masses
appear.
All parameters are correlated and compared with the nonrelativistic %%@
choices\cite{17,19,20} The quark masses,were  treated as fitting
parameters in a limited sense. Suitable values, allowed within the
experimental uncertainty in current quark masses \cite{1}, were
selected.
That parameterization is expected to lead to a reasonable reproduction of %%@
meson masses. No additional fitting was allowed when widths were
calculated.
  In that way one can reproduce
the measured $\Gamma (2 \gamma)$ reasonably well and make the
prediction for the $f_{0}(1370)$ $2 \gamma$ decay.

All results are based on the valent quark $q\overline{q}$
structure, which is characteristic of the model. The importance of
$q\overline{q}$ structure has often been mentioned
\cite{7}$-$\cite{14}. Our basic loop diagrams, Fig. 1 below,
correspond closely to the quark loop diagrams shown in Fig. 1 of
Ref.(10). Thus it is not surprising that our results, following
from the somewhat more complicated diagrams, depends strongly on
quark masses.

Moreover our model contains the sea component constrained by the
requirement of general Lorentz covariance, valid in an frame
\cite{16,18}. In that way, indirectly, some other QCD structures,
as discussed earlier \cite{16,18}, enter into our description.

Predictions based on our simplified model version test how well,
or how badly the model mimicks the real QED/QCD world.
The quarkonium approximation \cite{17} is investigated here in the %%@
circumstances which are different from the usual form factors
related problems\cite{16,17}.

The results depend also on the quark flavor structure of the
scalar mesons. They do distinguish among various propose
$u\overline{u}, d\overline{d}$ and $s\overline{s}$
mixing\cite{13,14} in $f_{0}$ mesons.We investigate only lower laying states %%@
without entering into discussion of the states as $a_0(1450)$
which would require to include the higher order terms of our
model.

\newpage

\vspace*{1pt}
\section{Brief description of the model}
\vspace*{-0.5pt} \setcounter{section}{2}
\renewcommand{\theequation}{\arabic{section}.\arabic{equation}}
\noindent A scalar meson $H$ with the four-momentum $P$ and  the
mass $M$ is covariantly represented \cite{16,18} by

\begin{displaymath}
|H(E,\vec{P}, M) \big> = \frac {N(\vec{P})}{(2\pi)^{3}}
\sum_{c,s_{1},s_{2}} \ \ \sum_{f} C_{f} \ \ \int  \ \big[
4m_{1}m_{2} \big] \ d^{4}p \ \delta (p^{2}-m^{2}_{1}) \ \Theta (e)
\end{displaymath}

\begin{equation}
\cdot d^{4}q  \ \delta(q^{2}-m^{2}_{2})  \ \Theta (\epsilon)  \
d^{4}K  \ F(K)\ \delta^{(4)} (p+q+K-P)  \ \Theta (E)  \ \phi_{f}
(l_{\perp})
\end{equation}

\begin{displaymath}
\cdot \overline{u}_{f,s_{1}}^{c} (\vec{p})
 v^{c}_{f,s_{2}} (\vec{q}) \ \
d^{+}_{f} (\vec{q}, c, s_{2}) b^{+}_{f} (\vec{p}, c, s_{1}) | 0
\big>
\end{displaymath}
Here $m_{i}$ are quark masses and $f$ stands for quark flavor. The
quark wave function is

\begin{equation}
\phi_{f}(l^{\mu}_{\bot}) = \frac{1} {(1 -
\frac{l_{\bot}^{2}}{4\beta_{f,H}^{2}})^{2}}
\end{equation}

\begin{displaymath}
l^{\mu}_{\bot} (P) = l^{\mu} - \frac{P^{\mu}(P \cdot l)}{M^{2}}
\end{displaymath}
with $l^{\mu} = (p -q)^{\mu}/2$, $p^{\mu}=(e,\vec{p})$ and
$q^{\mu}=(\epsilon,\vec{q})$. The fitting parameter $\beta_{f,H} \
\  (f=u,d,s,c)$ is fixed by fitting the meson mass as described
below. The dipole form (2.2) was found to be a better choice than
the exponential form used earlier \cite{16,18}. (See also some
remarks in Appendix.) Coefficient $C_{f}$ indicates the flavor
content of a particular meson (For example, see below formula
(3.14), where for $f_{0}(980) \ \ C_{u}=cos(\theta)/ \sqrt{2}$).

The symbols $\overline{u}, v, d^{+}, b^{+}$ correspond to valence
quarks while the sea function has a general form

\begin{equation}
F(K) = \delta^{(4)} \big[K^{\mu} - \frac{P^{\mu}}{M}
\frac{P^{\nu}\big(P- p - q\big)_{\nu}}{M} \big] \ \varphi(K).
\end{equation}
For simplicity we set

\begin{equation}
\varphi(K)=1.
\end{equation}
The complex looking Dirac function in (2.3) simplifies the model
structure easing all formal manipulations. One could produce a
somewhat more complicated model, without that Dirac function.
Additional parameter(s) in the sea model function would lead to a
richer and more flexible model \cite{18}. Thus the choice (2.4)
correspond to a minimalistic model version.

The scalar meson state (2.1) is normalized so that the matrix
element of the vector current $V^{\mu}$ would be, for example,

\begin{equation}
\langle H (\vec{P}_{f})| V^{\mu} | H (\vec{P}_{i})\rangle =
\frac{1}{(2 \pi)^{3}} F_{+}(Q^{2}) (P_{f} +P_{i})^{\mu} + \cdots
\end{equation}
with $Q^{\mu}=(P_{f} - P_{i})^{\mu}$, $F_{+}(Q^{2}=0) =1$. The
normalization (2.5) insures that the vector current, and thus
charge, is conserved. This requirement is equivalent to the
condition

\begin{displaymath}
\langle H (E,\vec{P},M)| H (E,\vec{P},M)\rangle = 2E =
\end{displaymath}

\begin{displaymath}
=\frac {N(\vec{P})^{2}}{(2\pi)^{6}} \sum_{c,s_{1},s_{2}}
 \sum_{ c',s'_{1},s'_{2}} \sum_{f,f'} C_{f} C_{f'}
\cdot \int d^{3}p \frac{m_{1}m_{2}}{e}  \frac{M}{E} \frac{\phi_{f}
(l_{\bot})}{q_{\parallel}} d^{3}p' \frac{m_{1}m_{2}}{e'}
\frac{M}{E} \frac{\phi_{f'}(l'_{\bot})}{q'_{\parallel}}
\end{displaymath}

$$
 \cdot[ \overline{
v}_{f',c',s'_{2}}(\vec{q'})
 u_{f',c',s'_{1}} (\vec{p'})
\overline{u}_{f,c,s_{1}} (\vec{p}) v_{f,c,s_{2}} (\vec{q})]\cdot
$$

\begin{equation}
\cdot \big<0| b_{f'} (\vec{p'}, c', s'_{1})  d_{f'} (\vec{q'}, c',
s'_{2}) d^{+}_{f} (\vec{q}, c, s_{2}) b^{+}_{f}(\vec{p}, c, s_{1})
| 0 \big> \big|_{ \vec{q}= T_1 , \ \ \vec{q'}= T_1' }
\end{equation}

Here $$T_1=- \vec{p} + \frac{\vec{P}}{M} (p_{\parallel})T$$
\begin{equation}
T=1+ \frac{ \sqrt{m^{2}_{2}-m^{2}_{1}+p^{2}_{\parallel}} }
{p_{\parallel}} \ \ \ ;\ \ \ \ \ p_{\parallel}= \frac{P \cdot
p}{M} \ \ \ ;\ \ \ \ \ q_{\parallel}= \frac{P \cdot q}{M}
\end{equation}
After a lengthy but straightforward manipulation, one find

\begin{displaymath}
N(\vec{P}) = \frac{E}{M} N(0)
\end{displaymath}

\begin{equation}
\langle H (E,\vec{P},M)| H (E,\vec{P},M)\rangle = 2E =3N(0)^{2} \sum_{f} %%@
C_{f}^{2} \int d^{3}p \ \frac{\epsilon}{e} \
{\big(}\frac{\phi_{f} (l_{\bot})}{q_{\parallel}}{\big)}^{2} \  (pq %%@
-m_{1}m_{2})
\end{equation}
>From (2.8) $N(0)$ can be calculated numerically.

The matrix element of the conserved vector current $V^{\mu}$ has
to vanish when current acts on the scalar meson state, i.e.

\begin{equation}
\langle 0| V^{\mu}  | H (E, \vec{P},M) \rangle \equiv 0
\end{equation}
The model states (2.1) are consistent with this very general
requirement. Some additional details are shown in the Appendix.

So far the model is closely related to ISGW model \cite{17}. In
the nonrelativistic limit and in the weak binding approximation it
goes exactly in the ISGW form\cite{16,18}. However the weak
binding approximation means that the quark masses and the quark
energies are approximately equal \cite{17}. In the present
  application, the model quark fields enter a loop (Fig. 1)
which constitutes the lowest QED approximation. The QCD
corrections are modeled by the functions (2.2) and (2.3). One can
try to mimick the real QED/QCD world by retaining small (current)
quark masses in the model. Then the meson mass should be equal to
a sum (weighted by the sea influence) of average model quark
energies. In the  model
  determined by (2.2)
and (2.4) this i

\begin{equation}
M= \frac {3 N(0)^{2}}{2E} \sum_{f} C_{f}^{2}
 \int d^{3}p \
(\frac{\phi_{f} (l_{\bot})}{q_{\parallel}})^{2} \ \frac{p \cdot q
- m_{1} m_{2}}{e/ \epsilon} (p_{\parallel} + q_{\parallel})|_{
\vec{q}= - \vec{p} + \frac{\vec{P}}{M} (p_{\parallel})T}
\end{equation}
As discussed in the Appendix this simple form holds in the
minimalistic
model version (2.4). The wave function $\phi_p$ can be connected with the %%@
usual potential$^{17,19,20}$ as shown in Appendix.  In the nonrelativistic, %%@
weak binding limit (WBL) (2.10) goes into

\begin{equation}
M \cong
\langle (e+\epsilon) \rangle \stackrel{WBL}{\rightarrow %%@
}\hat{m}_{1}+\hat{m}_{2}
\end{equation}
Here $\hat{m}_{i}$ are constituent quark masses \cite{17}. This WBL makes %%@
sense only if
one uses constituent masses $\hat{m}_{i}$, with the corresponding $\beta $'s %%@
\cite{17} in all relevant formulae.

If (2.10) is calculated explicitly in our model it can hold only for %%@
particular
values of model parameters, i.e. $\beta_{f,H}$ with a particular set of quark
%%@
masses. When one aims for $m_{i}$ close to the current quark
masses, that requires  the consistent
$\beta_{f,H}$  values. As explained in Appendix, by using Eq.(2.10), one %%@
determines the theoretical form factor at the
 physical momentum transfer $Q^2$. However, one is still dealing with some %%@
sort of a mock meson description.

Small quark masses,  which are used with the relativistic model, lead to the %%@
$\beta$ values which are close to those used in the nonrelativistic model. %%@
Full  comparisment  between those cases is given in Appendix.

\newpage
\section{Electromagnetic widths}
\setcounter{equation}{0} \setcounter{section}{3}
\renewcommand{\theequation}{\arabic{section}.\arabic{equation}}
\noindent In our model \cite{16,18} the amplitude ${\cal M}$ for
the transition $f_{0} \rightarrow 2\gamma$ is determined from the
leading diagrams shown in Fig. 1.

\begin{figure}[ht]
\begin{center}
\includegraphics[width=8.5cm,height=4.78cm]{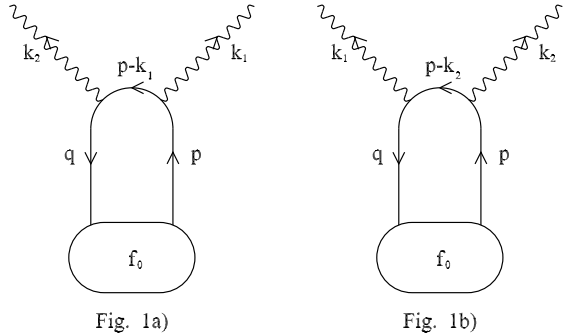}
\caption{Two-photon decay. Full lines are valence quarks, wavy
lines are emitted photons and blobs symbolize scalar meson
states.}
\end{center}
\end{figure}

Although the diagrams in Fig. 1 closely resemble the free quark
diagrams, they are not the same. One has to sum over the moment
$\vec{p}$, $\vec{q}$ and over the spins of the valence quark
states. These sums are weighted by the corresponding functions in
the meson state (2.1). A loop corresponds to each quark flavor.
For example, for the flavor $d$ the amplitude corresponding to the
diagram in Fig. 1a is determined by
\newpage

\begin{displaymath}
M_{1}^{\mu \nu} = (2\pi)^{3/2} \langle 0|:\overline {\Psi_{d}}
\gamma^{\mu} S_{F}(l_{2} - k_{1})\gamma^{\nu} \Psi_{d}:|(d
\overline {d})\rangle =
\end{displaymath}

\begin{displaymath}
= (2\pi)^{3/2} \sum_{c_{1},\alpha,\beta} \langle 0|: \int \frac
{d^{3}l_{1}}{(2\pi)^{3}} \frac {m_{d}}{l_{f}^{0}} \big[
b_{d}^{+}(\vec{l_{1}},\alpha,c_{1}) \overline
{u_{d}}(\vec{l_{1}},\alpha) +
d_{d}(\vec{l_{1}},\alpha,c_{1})\overline
{v_{d}}(\vec{l_{1}},\alpha) \big]
\end{displaymath}

\begin{equation}
\cdot \gamma^{\mu}
\frac{\not \! l_{2} - \not \! k_{1} +m_{d}}{(\not \! l_{2} - \not \! %%@
k_{1})^{2} - m_{d}^{2}} \gamma^{\nu}
\end{equation}

\begin{displaymath}
\cdot \int \frac {d^{3}l_{2}}{(2\pi)^{3}} \frac {m_{d}}{l_{2}^{0}}
\big[ b_{d}(\vec{l_{2}},\beta,c_{1})u_{d}(\vec{l_{2}},\beta) +
d^{+}_{d}(\vec{l_{2}},\beta,c_{1})v_{d}(\vec{l_{2}},\beta) \big]:
\end{displaymath}

\begin{displaymath}
\cdot \frac {N(\vec{P})}{(2\pi)^{3}} \sum_{c, s_{1},s_{2}} \int
d^{3}p \frac{m_{d}^{2}}{e}  \frac{M}{E} \frac{\phi_{d, f_{0}}
(l_{\bot})}{q_{\parallel}}
 \overline{u}_{d,c,s_{1}} (\vec{p}) v_{d,c,s_{2}} (\vec{q})
d^{+}_{d} (\vec{q}, c, s_{2}) b^{+}_{d}(\vec{p}, c, s_{1}) | 0
\big> \big|_{ \vec{q}= T_1 }
\end{displaymath}
The contraction of the creation  (annihilation) operators in (3.1)
leads to the summation over spin indicies. That give

\begin{displaymath}
M_{1}^{\mu\nu} = -3 \frac{N(0)}{(2\pi)^{3/2}}     \int d^{3}p
\frac{m_{d}^{2}}{e}  \frac{\phi_{d, f_{0}}
(l_{\bot})}{q_{\parallel}}
\end{displaymath}

\begin{equation}
\cdot Tr [\gamma^{\mu} \frac{\not \! p - \not \! k_{1} +m_{d}}{(-2
p k_{1})} \gamma^{\nu} \frac{\not \! p  +m_{d}}{2 m_{d}}
\frac{\not \! q  - m_{d}}{2 m_{d}}] \big|_{ \vec{q}= - \vec{p} +
\frac{\vec{P}}{M} (p_{\parallel})T } \equiv I(\vec{p},\vec{q})
Tr(k_{1})^{\mu\nu}
\end{equation}
The amplitude corresponding to Fig. 1 is

\begin{equation}
{\cal M}= M_{1}^{\mu\nu}
\epsilon_{\mu}(k_{2},\lambda_{2})\epsilon_{\nu}(k_{1},\lambda_{1})
+ M_{2}^{\mu\nu}
\epsilon_{\mu}(k_{1},\lambda_{1})\epsilon_{\nu}(k_{2},\lambda_{2}),
\\
\end{equation}

\begin{displaymath}
M_{2}^{\mu\nu}=I(\vec{p},\vec{q}) Tr(k_{1} \rightarrow
k_{2})^{\mu\nu}
\end{displaymath}
Routine calculation of traces produces the final result which
contains

\begin{equation}
Tr(k_{1})^{\mu\nu}=A_{1}(2p^{\nu}q^{\mu} - 2 p^{\mu} p^{\nu} +
k_{1}^{\mu} (p-q)^{\nu} + k_{1}^{\nu} (p-q)_{\mu} - g^{\mu\nu}
\big[k_{1} (p-q)\big])
\end{equation}

\begin{displaymath}
A_{1,2}= \frac{1}{2m_{d}(p \cdot k_{1,2})}
\end{displaymath}
It is convinient to carry out further calculation in the meson rest frame %%@
(MRF). That is defined by

\begin{displaymath}
P^{\mu}= (M, \vec{0}) \ ; \ \ \ e=\epsilon \ ; \ \ \ \vec{q}= -
\vec{p}
\end{displaymath}

\begin{equation}
k_{1}^{\mu} = (\omega, \vec{k}) \ ; \ \ \ k_{2}^{\mu} = (\omega, -\vec{k}) \ ;
%%@
\
\ \ \omega = |\vec{k}| = \frac {M}{2}
\end{equation}
Further simplification is obtained by selecting orthogonal polarization %%@
vectors

\begin{displaymath}
P \cdot \epsilon_{1,2} = k_{1} \cdot \epsilon_{1} = k_{2} \cdot \epsilon_{2} =
%%@
0
\end{displaymath}

\begin{equation}
\epsilon^{\mu}(k_{2},\lambda_{2}) \equiv \epsilon^{\mu}_{2} = (0,
\vec{\epsilon}_{2}) \ ; \ \  \ \epsilon^{\mu}(k_{1},\lambda_{1})
\equiv \epsilon^{\mu}_{1} = (0, \vec{\epsilon}_{1}) \ ; \ \  \
k_{2} \cdot \epsilon_{1} = k_{1} \cdot \epsilon_{2} = 0
\end{equation}
As shown in the next section the result does not depend on a
particular gauge. With (3.5), (3.6) one obtain

\begin{equation}
{\cal M}= -2I(\vec{p},\vec{q})\{ A_{1}
\big[(\vec{\epsilon}_{1}\vec{\epsilon}_{2}) (\vec{k} \vec{p}) +
2(\vec{p}\vec{\epsilon}_{1}) (\vec{p}\vec{\epsilon}_{2})\big] +
A_{2}\big[-(\vec{\epsilon}_{1}\vec{\epsilon}_{2}) (\vec{k}
\vec{p}) + 2(\vec{p}\vec{\epsilon}_{1})
(\vec{p}\vec{\epsilon}_{2})\big] \}
\end{equation}

\begin{displaymath}
I(\vec{p},\vec{q})|_{\vec{P}=0} = -3(2\pi)^{3/2} N'(0) \int d^{3}p
\frac{m_{d}^{2}}{e^{2}} \frac{1}{(1 +
\frac{\vec{p}^{2}}{4\beta_{d,f_{0}}^{2}})^{2}}
\end{displaymath}
By using $\vec{p}\vec{k} = p \omega cos\theta$ one obtains

\begin{displaymath}
{\cal M}= - \frac{3 N(0)}{m_{d}\omega (2\pi)^{1/2}} \int p^{2}dp \
sin\theta d \theta \ \frac{m_{d}^{2}}{e^{2}} \frac{1}{(1 +
p^{2}/(4\beta_{d,f_{0}}^{2}))^{2}}
(\vec{\epsilon}_{1}\vec{\epsilon}_{2}) \frac{2 p^{2} (\omega
cos^{2}\theta + e sin^{2}\theta)}{e^{2} - p^{2} cos^{2}\theta }
\end{displaymath}

\begin{equation}
\equiv
 (\vec{\epsilon}_{1}\vec{\epsilon}_{2}) \cdot I_{d,\overline{d}}(f_{0})
\end{equation}
The calculation of the decay width

\begin{equation}
\Gamma = \frac{1}{32 \pi M_{H}} \overline{ |{\cal M}_{H}|^{2} }
\end{equation}
requires the summation over photon polarization states,

\begin{equation}
\overline{\sum_{\lambda_{1}\lambda_{2}} |{\cal M}|^{2} } = 2 \cdot
I_{f,\overline{f}}(H)^{2}
\end{equation}
as well as the summation over quark flavors in (3.8), connected
with the meson quark structure, which we parameterize a $|H\rangle
= \sum_{f} C_{f} |f,\overline{f}\rangle$.

\begin{displaymath}
|f_{0}(980)\rangle = \frac{cos\theta}{\sqrt{2}}(|u \overline{u}\rangle + |d %%@
\overline{d}\rangle) + sin\theta |s \overline{s}\rangle
\end{displaymath}

\begin{displaymath}
|a_{0}(980)\rangle = \frac{1}{\sqrt{2}}(|u \overline{u}\rangle - |d %%@
\overline{d}\rangle)
\end{displaymath}

\begin{displaymath}
|f_{0}(1370)\rangle = \frac{-sin\theta}{\sqrt{2}}(|u \overline{u}\rangle + |d
%%@
\overline{d}\rangle) + cos\theta |s \overline{s}\rangle
\end{displaymath}

\begin{equation}
|\chi_{c_{0}}(3415)\rangle = |c \overline{c}\rangle
\end{equation}
The physical mixing for these states is usually determined
\cite{14} by

\begin{equation}
cos\theta = 1 \ ; \ \ \  sin\theta = 0
\end{equation}
or

\begin{equation}
cos\theta = \frac{1}{3} \ ; \ \ \  sin\theta = \frac{2\sqrt{2}}{3}
\end{equation}
Eventually one finds by summing over flavors:

\begin{displaymath}
Int(f_{0}(980)) =\frac{e^{2}}{9} [\frac{5cos\theta}{\sqrt{2}} I_{u
\overline{u}}(f_{0}(980)) + sin\theta I_{s
\overline{s}}(f_{0}(980))]
\end{displaymath}

\begin{displaymath}
Int(a_{0}(980)) =\frac{e^{2}}{3\sqrt{2}} I_{u
\overline{u}}(a_{0}(980))
\end{displaymath}

\begin{displaymath}
Int(f_{0}(1370)) =\frac{e^{2}}{9} [\frac{-5sin\theta}{\sqrt{2}}
I_{u \overline{u}}(f_{0}(1370)) + cos\theta I_{s
\overline{s}}(f_{0}(1370))]
\end{displaymath}

\begin{equation}
Int(\chi_{c_{0}}(3415)) = \frac{4e^{2}}{9}I_{c %%@
\overline{c}}(\chi_{c_{0}}(3415))
\end{equation}
where $I_{f,\overline{f}}(H)$ is given by formula (3.8) and $e$ is
the electron unit charge , i.e. $e^{2}/(4\pi) = \alpha =
(137,04)^{-1}$.

The $2\gamma$-decay width can be put in the final form valid for a
meson $H$ from (3.11):

\begin{equation}
\Gamma(H) = \frac{\pi \alpha^{2}}{M_{H}}
(\frac{Int(H)}{e^{2}})^{2}
\end{equation}

\newpage
\section{Gauge invariance explicitly tested}
\setcounter{equation}{0} \setcounter{section}{4}
\renewcommand{\theequation}{\arabic{section}.\arabic{equation}}
\noindent Through its covariant nature, being in a sense a
simplified rendering of the real QCD field theory, the model
\cite{16,18} automatically produces gauge invariant results.

Under the gauge transformation

\begin{equation}
\epsilon_{i \mu} \rightarrow \epsilon_{i \mu} + \Lambda k_{i \mu}
\end{equation}
the amplitude (3.3) should not change. That leads to the equality

\begin{displaymath}
Tr(k_{1})^{\mu\nu} (\epsilon_{2} + \Lambda k_{2})_{\mu}
(\epsilon_{1} + \Lambda k_{1})_{\nu} + Tr(k_{1} \rightarrow
k_{2})^{\mu\nu} (\epsilon_{1} + \Lambda k_{1})_{\mu} (\epsilon_{2}
+ \Lambda k_{2})_{\nu} =
\end{displaymath}

\begin{equation}
= Tr(k_{1})^{\mu\nu} \epsilon_{2 \mu} \epsilon_{1 \nu} + Tr(k_{1}
\rightarrow k_{2})^{\mu\nu} \epsilon_{1 \mu} \epsilon_{2 \nu} +
N(\Lambda, k_{1}, k_{2})
\end{equation}
Here first two terms give the amplitude (3.3). The piece
$N(\Lambda, k_{1}, k_{2})$ should not contribute to the physical
amplitude (3.7). In the meson rest frame, using (3.5) and (3.6),
one immediately finds

\begin{equation}
N(\Lambda, k_{1}, k_{2}) = \Lambda \frac{2\omega}{m_{d}}
(\frac{\vec{p} \vec{\epsilon_{1}}}{p \cdot k_{1}} + \frac{\vec{p}
\vec{\epsilon_{2}}}{p \cdot k_{2}}) [\omega - e(\vec{p})]
\end{equation}
while the terms proportional to $\Lambda^{2}$ cancel. $N(\Lambda,
k_{1}, k_{2})$ enters the integration over the bound quark
momentum $\vec{p}$, as shown in (3.2), (3.3). The corresponding
change in ${\cal M}$ is

\begin{equation}
\Delta{\cal M} = I(\vec{p}, \vec{q})|_{\vec{P}=0} \cdot N(\Lambda,
k_{1}, k_{2}).
\end{equation}
Here one has

\begin{displaymath}
\int d^{3}p = \int_{0}^{\infty} p^{2}dp \int_{0}^{\pi} sin\theta
d\theta \int_{0}^{2\pi} d\Phi
\end{displaymath}

\begin{equation}
\vec{p} \vec{\epsilon_{1,2}} =|\vec{p}| [ sin\theta cos\Phi
(\hat{x}\vec{\epsilon_{1,2}}) + sin\theta sin\Phi
(\hat{y}\vec{\epsilon_{1,2}}) ]
\end{equation}
Integration over the azimuthal angle $\Phi$ gives zero result, so
one has

\begin{equation}
\Delta{\cal M} \equiv 0
\end{equation}
as required by the gauge invariance.

\newpage
\section{Results and discussion}
\setcounter{equation}{0} \setcounter{section}{5}
\renewcommand{\theequation}{\arabic{section}.\arabic{equation}}
\noindent The application of the model (2.1) starts with the
self-consistency condition (SCC) (2.10). Quark masses are
selected, as close as practical to the Particle Data \cite{1}
values. Than parameters $\beta_{f,H}$ are varied for various
flavors appearing in (3.11) so as to reproduce the experimental
meson masses \cite{1}.

\begin{displaymath}
M[f_{0}(980)] = 0.980 \ GeV \ ;\ \ \ \ \ \ M[a_{0}(980)] = 0.980 \
GeV
\end{displaymath}

\begin{equation}
M[f_{0}(1370)] = 1.370\ GeV \ ;\ \ \ \ \ \ M[\chi_{c_{0}}(3415)] =
3.415 \ GeV
\end{equation}

Theoretical expression (2.10), (2.11) is a sum of parts
corresponding to various flavors. For example

\begin{equation}
M[a_{0}(980)]_{th} = \frac{1}{2} (M_{u} +M_{d}).
\end{equation}

It turns out that SCC requires light quark masses somewhat larger
than Particle Data \cite{1} median values, while strange and charm
masses could be kept within Particle Data limit. The satisfactory
selection i

\begin{equation}
m_{u} = m_{d} = 0.015 \ GeV \;\ \ \ m_{s} = 0.120 \ GeV \;\ \ \
m_{c} = 1.5 \ GeV
\end{equation}
The most stringent restriction on the light quark parameters is
obtained by fitting the mass of the $a_{0}$ meson. If there is no
strangeness mixing, i.e. with $\theta = 0$ in (3.11), the same
mass is calculated for the $f_{0}(980)$ meson too. Some
interesting values are shown in Table 1.

\begin{table}[h]
\caption{Mock mass values for $a_{0}$}
\begin{center}
\begin{tabular}{|c|c|}
\hline $\beta_{u,d} (GeV)$ & $M_{0} (GeV)$  \\ \hline 0.260 &
0.885 \\ 0.270 & 0.919 \\ 0.280 & 0.953 \\ 0.288 & 0.980 \\ 0.295
& 1.004 \\ \hline
\end{tabular}
\end{center}
\end{table}

With ideal  mixing $(\theta = 0)$ $f_{0}(1370)$ has a pure $s
\overline{s}$ configuration. The corresponding mass values are
shown in Table 2.

\begin{table}[h]
\caption{Mock mass values for $s \overline{s}$ configuration}
\begin{center}
\begin{tabular}{|c|c|}
\hline $\beta_{s} (GeV)$ & $M_{0} (GeV)$  \\ \hline 0.350 & 1.272
\\ 0.381 & 1.370 \\ 0.400 & 1.435 \\ 0.450 & 1.599 \\ 0.500 &
1.764 \\ \hline
\end{tabular}
\end{center}
\end{table}

The mass of the pure $c \overline{c}$ state $\chi_{c_{0}}$ can be
reproduced by using $\beta_{c} = 0.267 \ GeV$ as shown in Table 3.

\begin{table}[h]
\caption{Mock mass values for $c \overline{c}$ configuration}
\begin{center}
\begin{tabular}{|c|c|}
\hline $\beta_{c} (GeV)$ & $M_{0} (GeV)$  \\ \hline 0.250 & 3.377
\\ 0.267 & 3.415 \\ 0.280 & 3.445 \\ 0.300 & 3.490 \\ 0.330 &
3.560 \\ \hline
\end{tabular}
\end{center}
\end{table}

When non\-ideal  mixing (3.11), (3.13) is allowed the masses of
$f_{0}(980)$ and $f_{0}(1370)$ can be reproduced by $\beta_{u}$
and $\beta_{s}$ which are different from those shown in Tables 1
and 2. However, the values in Table 1 still correspond to the
$a_{0}$ mass. The masses $M[f_{0}(980)]$ and $M[f_{0}(1370)]$ are
reproduced by:

\begin{equation}
\beta_{u,d} = 0.419 \ GeV \ ;\ \ \ \ \ \  \beta_{s} = 0.242 \ GeV
\end{equation}
The corresponding $\Gamma(H \rightarrow 2\gamma)$ values are
summarized in Table 4.

\begin{table}[h]
\caption{Decay widths}
\begin{center}
\begin{tabular}{|c|c|c|c|}
\hline Meson & Mixing & $\Gamma_{theory}  (keV)$ & $\Gamma_{exp}
(keV)$ \\ \hline $a_{0}(980)$ &        & 0.137 & $0.26 \pm 0.08 $
\\ $f_{0}(980)$ & (3.12) & 0.380 & $0.56 \pm 0.11$ \\ $f_{0}(980)$
& (3.13) & 0.534 &  $0.56 \pm 0.11$ \\ $f_{0}(1370)$ & (3.12) &
0.348 &     \\ $f_{0}(1370)$ & (3.13) & 0.145 &      \\
$\chi_{c_{0}}(3415)$ &    & 4.608 & $4.0  \pm 2.8$ \\ \hline
\end{tabular}
\end{center}
\end{table}

All conclusions depend strongly on the quark masses. For example,
if one chooses $m_{c} = 1.4\ GeV$ than the SCC  requires
$\beta_{c} = 0.346 \ GeV$.

The $q \bar q$ structure, which is the main feature of the model,
might be capable of explaining two photon decay. By that one does
not mean a naive "free-quark" structure of an early
nonrelativistic model. In the present model the valence quarks are
immersed in a sea, which, however rudimentary, takes into the
account the interference
of other QCD induced configurations like for example $s\bar{s}$ pairs, gluons
%%@
etc. The SCC (2.10) transmits that into the numerical results.

The experimental error in $f_{0}(980) \to 2 \gamma$ rate is rather
large. Although, the large theoretical prediction in Table 4,
seems to be in better agreement with experiments, the smaller one,
which corresponds to the ideal mixing, cannot be ruled out.
However, the $f_0(1370)$ decay into pions indicates the presence
of the light $q \bar q$ combinations\cite{7,10,11}. Our result
also agrees with the nonideal mixing as considered by
Lanik\cite{13}. If the corresponding theoretical predictions
(Table 4) for the decay of $f_{0} (1370)$ turns out to be at least
approximately correct, one would have a very strong support for
nonideal mixing\cite{13,14}.

The experimental data \cite{1} for the $a_{0}(980) \to 2 \gamma$
decay width  contain large errors. Our theoretical value, Table 4,
is close to the lower experimental limit. Various other
theoretical approaches are summarized in Ref.(9).  Our approach
has some analogy with Deakin et al.\cite{10} who used constituent
quark masses and concluded that theoretical results depend
strongly on the numerical values of those masses. The same strong
dependence on the masses, was found here. The decay width and the
mass of $\chi_{c_{0}}$ are very well reproduced within the model.

Naturally all our conclusions depend on the validity of model as
such. Here we have tried a simplicistic version of the model,
which relies on the functions (2.2) and (2.4) That gave SCC (2.10)
which represents a strong restriction on the model parameters.
By selecting other functions instead (2.2) and (2.4) one would end with less %%@
restrictive SCC.

The model which was employed here indicates the importance of the
valence $q \bar q$ structure \cite{17} in the meson state. There
is some hope that such relativistic model, at least in some richer
version, can play a useful role in the classification of meson
states. It can be a useful tool in the design of future
experiments if it can provide a reasonable estimate of the
magnitude of expected experimental effects.

\newpage
\begin{Large}
\appendix{\bf Appendix}
\end{Large}
\vspace{0.5cm} \setcounter{equation}{0}
\renewcommand{\theequation}{A\arabic{equation}}

\noindent If one evaluates the expression (2.10) for arbitrary
$\beta_{f}$'s, one obtains a value $M_{0}$, which is different
from the physical mass M.

Thus the model meson state (2.1) is an approximation of the real
physical state (of a meson with mass $M$ and momentum $P^{\mu}$)
in the sense that there is a one to one correspondence between
physical state with velocity $v^{\mu}  =P^{\mu}/M$ (with respect
to the meson rest frame) and model state with the same velocity
$v^{\mu}$. This can be seen explicitly from formulae (2.6) and
(2.7), where the frame dependence of the internal quark momenta is
described only through the velocity components $E/M$ and
$\vec{P}/M$ and/or through the Lorentz scalar quantities
\cite{16,18}.

$M_{0}$  has some similarity with so called  "mock mass"
\cite{17}. Therefore, a model state (2.1) correspond also to the
different momentum $P^{\mu}_{0}$ given by

\begin{equation}
P^{\mu}_{0} = v^{\mu} \cdot M_{0} = \frac{P^{\mu}}{M} \cdot M_{0}
.
\end{equation}

Consequently, when calculating a physical quantity dependent on a
square of the physical momentum transfer $Q^{2}$, one obtains the
value of that quantity at the momentum transfer $Q_{0}^{2}$ which
is shifted by the factor $M_{0}^{2}/M^{2}$ to the physical one,

\begin{equation}
Q_{0}^{2} = \frac{M_{0}^{2}}{M^{2}} \cdot Q^{2}.
\end{equation}

This shift is especially important in processes described by the
hadronic matrix element of the form $\langle 0| \Gamma^{\mu} |A
\rangle$, as for example in leptonic meson decays or $A
\rightarrow \gamma \gamma$ transitions etc. If $M_{0}^{2} \neq
M^{2}$ one obtains amplitude $T(M_{0}^{2} )$ instead of $T(M^{2}
)$. For light mesons $(\pi, K...)$
approximation of $T(M^{2} )$ by $T(M_{0}^{2} )$ is poor. Thus, Ref.(20) has %%@
introduced suitable corrections.

For heavy mesons such approximation is much better and so it was
not even mentioned in our previous work \cite{16,18}.

In the nonrelativistic quark model \cite{17} "mock-mass" was
defined simply as a sum of the constituent quark masses. A formal,
covariant  expression for that is the expectation value of the
valence quark (antiquark) momentum operators

\begin{displaymath}
\hat{k}^{\mu}_{f} = \int \frac{d^{3}k}{(2 \pi)^{3}}
\frac{m_{f}}{k^{0}} \  k^{\mu} \ \sum_{s} \ \big[
b^{+}_{f}(\vec{k}, s) b_{f}(\vec{k}, s) + d^{+}_{f}(\vec{k}, s)
d_{f}(\vec{k}, s) \big] \  .
\end{displaymath}

\begin{equation}
\hat{P}_{0}^{\mu} = \ \sum_{f} \hat{k}^{\mu}_{f}
\end{equation}

One obtain

\begin{equation}
 \frac {1}{2E} \langle H(E, \vec{P},M)| \hat{P}^{\mu}_{0}  | H (E, %%@
\vec{P},M)\rangle |_{ \beta \neq \beta_{H}} =  \frac{P^{\mu}}{M}
\cdot M_{0}
\end{equation}

\begin{displaymath}
= \frac {3 N(0)^{2}}{2E} \sum_{f} C_{f}^{2}
 \int d^{3}p \
(\frac{\phi_{f} (l_{\bot})}{q_{\parallel}})^{2} \ \frac{p \cdot q
- m_{1} m_{2}}{e / \epsilon} (p^{\mu} + q^{\mu})|_{ \vec{q}= -
\vec{p} + \frac{\vec{P}}{M} (p_{\parallel})T}
\end{displaymath}

\begin{displaymath}
= \frac {P^{\mu}}{M}  \cdot \frac {3 N(0)^{2}}{2E} \sum_{f}
C_{f}^{2}
 \int d^{3}p \
(\frac{\phi_{f} (l_{\bot})}{q_{\parallel}})^{2} \ \frac{p \cdot q
- m_{1} m_{2}}{e/ \epsilon} ( p_{\parallel} +q_{\parallel} ) |_{
\vec{q}= - \vec{p} + \frac{\vec{P}}{M} (p_{\parallel})T}
\end{displaymath}
Here $M_{0}$ is  a Lorentz scalar quantity which satisfies
$\langle \hat{P}^{\mu}\rangle^{2}=M_{0}^{2}$. However, this does not mean that
%%@
our state (2.1)
is really an eigenstate of the meson four-momentum. The operator %%@
$\hat{P}^{\mu}$, which contains only the free quark operators, is
a mock-meson operator. Our model is
Lorentz covariant but it is not relativistic in the quantum field theory %%@
sense. The model mocks a hyperplane projected solution of a
Bethe-Salpeter equation. It corresponds to a quasi potential
approximation.

For a "real" meson, one should have

\begin{equation}
\frac {1}{2E} \langle H(E, \vec{P},M)| \hat{P}_{0}^{\mu}  | H (E, %%@
\vec{P},M)\rangle |_{\beta=\beta_{f,H}} \equiv P^{\mu}
\end{equation}
Here, the quotation marks symbolize the pseudo realistic ( mock ) character of
%%@
a meson state.
In the rest frame $ P^{\mu} = (M,0,0,0)$ this determines a mock mass $M_{0}$,
%%@
which
is given by (2.10) ( $M\rightarrow M_{0}$ ). The expression (2.10) is a %%@
normalization integral
(2.8), multiplied by the factor $ p_{\parallel} +  q_{\parallel} $, which is,
%%@
in the rest frame,
 the sum of quark energies $e + \epsilon $.

 One reaches WBL by introducing large constituent masses $\hat{m}_i$ \cite{17}
%%@
and by going into
 nonrelativistic limit:

\begin{equation}
M_0= N_{WBL}^{2} \sum_{f} C_{f}^{2}
 \int d^{3}p \ (\vec{p})^{2} \phi_{f} (l_{\bot})^{2}
(\hat{m}_{1} + \hat{m}_{2})= \hat{m}_{1} + \hat{m}_{2}
\end{equation}

Working with a more general expression (A4) one can enforce $M_{0}
= M$ by appropriate choice of the model parameters $\beta$ and
$m_i$.
That choice of parameters ensures that one works at the kinematically correct
%%@
point $Q_0^2 \equiv Q^2$.

The present model parameter $\beta$ is dependent not only on the
quark-flavours, but also on the meson mass. Thus for example
$\beta_{u}$ has different values in $f_{0}(980)$ and $f_{0}(1370)$
mesons.

In the simplest version of the model \cite{16,18} considered here,
the role of the quark-gluon sea described by the momentum
$K^{\mu}$ is mostly kinematical. In principle the sea could enter
into (2.1)  dynamically also, affecting both, the internal
momentum distribution  $\phi$ and the internal spin distribution.
These possibilities, sketched in Ref.(18), are not explored here.
To some extent their effects were taken into account
phenomenologically by fitting the parameter $\beta_{f,H}$.

The model parameters can be also connected with the usual Coulomb plus linear
%%@
potential\cite{17,19,20}

\begin{equation}
V(r) =  - {\frac{4\,\alpha }{3\,r}}+ b\,r + c \label{a7}
\end{equation}

The comparison with the pseudoscalar meson applications\cite{17} will be %%@
facilitated if that case is briefly revised first. The relativistic case is %%@
described by the formulae of Ref.(23) in which the potential (\ref{a7}) must %%@
be used. Their formulae are pseudoscalar version of our expression (\ref{a8})
%%@
below.

The relativistic model fit requires slight readjustment of parameters. One has
%%@
to use $c=0$ in order to reproduce masses. However the $\beta$ values, which %%@
were found by variational procedure\cite{17} do not depend on c. In Fig. A.1.
%%@
the $\beta$'s for the relativistic and nonrelativistic\cite{17} fit are %%@
compared.

\begin{figure}[h]
\begin{center}
\includegraphics[width=8.5cm,height=5.18cm]{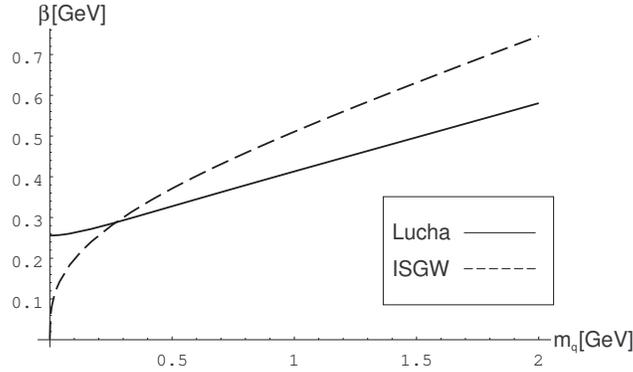}
\caption{ISGW\cite{17} and Lucha\cite{21} model $\beta$'s for pseudoscalar %%@
$|q\bar{q}\rangle$}
\end{center}
\label{fig2}
\end{figure}

In the relativistic approach one finds sizable $\beta$ values (comparable with
%%@
nonrealtivistic ones), for small quark masses also. For $q\bar{q}$ pairs the %%@
ratios among calculated masses are for example  %%@
$m_{NR}(u\bar{d})/m_{R}(u\bar{d})\cong 0.7$, %%@
$m_{NR}(u\bar{s})/m_{R}(u\bar{s})\cong 0.8$.

The scalar meson masses can be connected with (\ref{a7}) by

\begin{equation}
M_0=\frac{\sum_f C_f^2 (f(\beta,m_q)+h(\beta,m_q))}{\sum_f C_f^2 \ %%@
g(\beta,m_q)} \label{a8}
\end{equation}

\begin{equation}
f(\beta,m_q)=\int_0^{\infty} 4 \pi p^2 \, dp \frac{2 p^2 \phi_f^2
}{p^2+m_f^2}2\epsilon_f
\end{equation}
\begin{equation}
h(\beta,m_q)=\int_0^{\infty} 4 \pi p^2 \, dp \frac{2 p^2 \phi_f^2
}{p^2+m_f^2}V(p)
\end{equation}
\begin{equation}
g(\beta,m_q)=\int_0^{\infty} 4 \pi p^2 \, dp \frac{2 p^2 \phi_f^2
}{p^2+m_f^2}
\end{equation}

This is  just the expression (A.4) in meson rest frame with
(\ref{a7})
included and $g$ is normalization. With the potential (\ref{a7})$(c=0, %%@
\alpha=0.63)$ one finds for example for $a_0(980): \beta=0.284$ and $M_0=977\
%%@
MeV$. It is useful to note that one obtains $h=0$ for $\beta=0.288$ and $\sum
%%@
h/g\cong 0.011 \ MeV $ for $\beta=0.284$. Thus the expressions (A.4) and (A.8)
%%@
are numerically consistent for our range of parameters.

The model also satisfies a very general condition (2.9). Using

\begin{equation}
V^{\mu}(0) = \sum_{f} :\overline{\Psi}_{f}(0) \gamma^{\mu}
\Psi_{f}(0) :
\end{equation}
where

\begin{equation}
\Psi_{f}(0) = \sum_{s} \int \frac {d^{3}l}{(2\pi)^{3}} \frac
{m_{f}}{E_{f}} [ b_{f}(\vec{l},s)u_{f}(\vec{l},s) +
d^{+}_{f}(\vec{l},s)v_{f}(\vec{l},s) ]
\end{equation}
the condition (2.12) can be explicitly written as:

\begin{equation}
\langle 0| V^{\mu}(0) |f_{0}\rangle = \frac {1}{(2\pi)^{3/2}}
I(\vec{p},\vec{q}) \frac {m_{1}q^{\mu} - m_{2}p^{\mu}}
{m_{1}m_{2}}
\end{equation}
The conserved vector current requires equal quark masses, i.e
$m_{1} = m_{2}$. In the rest frame $E=M,\ \vec{P}=0$,
$\vec{p}=-\vec{q}$,  $\epsilon = e$. For the time component of
(A6) one obtains

\begin{equation}
 \frac {m_{1}q^{0} - m_{2}p^{0}}
{m_{1}m_{2}} = \frac {\epsilon - e}{m_{1}} = 0.
\end{equation}
The spatial components $\mu =1,2,3$ vanish after the integration
over $d^{3}p$ in $I(\vec{p},\vec{q})$ (3.7).

A more general scalar meson state than (2.1) can be constructed by
replacing

\begin{equation}
\overline{u}_{f}v_{f} \rightarrow \overline{u}_{f} (a + b \frac
{\not \! P }{M})v_{f}
\end{equation}
and by adjusting the normalization accordingly. It can be shown
that the term proportional to $b$ neither influences the
normalization, nor contributes to the decay width (3.15) within
conventions, such as for example (2.4), which were used here.
Therefore the  simpler form (2.1) was used.

\newpage
%\nonumsection{References}

\end{document}